\documentclass[a4paper]{article}
\usepackage[T1]{fontenc} 
\usepackage{float}       
\usepackage{helvet}      
\usepackage{mathptmx}    
\usepackage{rsc}         
\usepackage{setspace}    
\usepackage{graphicx, graphics}
\usepackage{appendix}
\usepackage{subeqnarray}
\usepackage{pifont}
\usepackage[latin1]{inputenc}

\AtBeginDocument{\doublespacing}
\newfloat{chart}{htbp}{loc}
\newfloat{graph}{htbp}{loh}
\newfloat{scheme}{htbp}{los}

\usepackage[version=3]{mhchem} 

\newcommand\Xsec{\mbox{sec}}
\newcommand\Xlog{\mbox{ln}}

\author{%
  Aleksey V. Belyaev
  \thanks{%
        Department of Physics, M. V. Lomonosov Moscow State University, 119991 Moscow, Russia
  }\,\,\thanks{A. N. Frumkin Institute of Physical Chemistry and Electrochemistry, Russian Academy of Sciences, 119991 Moscow, Russia}
  \and
  Olga I. Vinogradova \protect\footnotemark[2]\,\,\thanks{%
      Department of Chemistry, M. V. Lomonosov Moscow State University, 119991 Moscow, Russia
}%
\,\,\thanks{ITMC and DWI,  RWTH Aachen, Pauwelsstr. 8, 52056 Aachen, Germany}
  }

\title{Hydrodynamic interaction with super-hydrophobic surfaces}

\begin{document}

\maketitle

\begin{abstract}
Patterned surfaces with large effective slip lengths, such as super-hydrophobic surfaces containing trapped gas bubbles, have the potential to reduce hydrodynamic drag. Based on lubrication theory, we analyze an approach of a hydrophilic disk to such a surface. The drag force is predicted analytically and formulated in terms of a correction function to the Reynolds equation, which is shown to be the harmonic mean of
corrections expressed through effective slip lengths in the two principal (fastest and slowest) orthogonal directions. The reduction of drag is especially pronounced for a thin (compared to texture period) gap. It is not really sensitive to the pattern geometry, but depends strongly on the fraction of the gas phase and local slip length at the gas area.

\end{abstract}

\section{Introduction}

It is more than 100 years since Reynolds published his famous analysis of hydrodynamic lubrication due to a thin liquid film confined between two moving solids~\cite{reynolds1886}. This theory is based on the simplification of the Navier-Stokes equations of continuum hydrodynamics by exploiting the special geometry
of a film and no-slip boundary conditions. The utility of this theory still stands today and many extensions and applications of the original analysis maybe found in numerous publications.

Classical solutions of creeping flow equations of the lubrication theory for a circular disk of radius $R$ moving with a velocity $U$ towards a smooth wall, the so-called Reynolds problem, gives~\cite{reynolds1886}
\begin{equation}\label{F0}
  F_R= \frac{3}{2}\pi\eta U \frac{R^4}{H^3},\,\,\, {\rm provided} \,\,\, \frac{\rho H U}{\eta} \ll 1.
\end{equation}
when the gap $H$ becomes small compared to $R$. Here $\eta$ denotes the fluid viscosity. A consequence
of this lubrication effect is that the close approach a disk to the wall, or its pulling away from it would take large time. This is the basis of the phenomenon of viscous adhesion, used in adhesives such as `Scotch tape' or in the `wringing' together of smooth metal surfaces~\cite{batchelor.gk:2000}, but may represent a very unfavorable scenario for other applications. An efficient strategy for reducing the drag force is to exploit hydrodynamic slip, which can be generated at hydrophobic surfaces and is quantified by the slip
length $b$ (the distance within the solid at which the flow profile extrapolates to zero)~\cite{vinogradova1999,lauga2005,bocquet2007}. The near-field hydrodynamic interaction of a hydrophilic disk with such a hydrophobic surface (a situation which allows one to avoid a formation of a gas bridge in the gap~\cite{andrienko.d:2004}) leads to a correction to the Reynolds force~\cite{vinogradova.oi:1995d}

\begin{equation}\label{fast}
  f^{\ast} = \frac{F}{F_R} = \frac{H+b}{H+4 b}
\end{equation}
Depending on the ratio $b/H$, the correction for slippage $f^{\ast}$ can turn to $1$ (large compared to slip length distances) or $1/4$ (small distances). Since for hydrophobic smooth and homogeneous surfaces $b$ can be of the order of tens of nanometers ~\cite{vinogradova2003,charlaix.e:2005,joly.l:2006,vinogradova.oi:2009}, but not much more, it is impossible to benefit of such a nanometric slip at separations O($\mu$m) and larger.

Hydrophobicity can be significantly amplified
by roughness, and extreme hydrophobicity can be generated with
well-controlled textures~\cite{quere.d:2005}. Such super-hydrophobic (SH) surfaces in the Cassie state, i.e. where the texture is filled with gas,
  can reduce friction due to trapped nanobubbles~\cite{vinogradova.oi:1995b,cottin_bizonne.c:2003.a}, leading to a many-micron effective slip lengths~\cite{ou2005,joseph.p:2006,choi.ch:2006}. A mechanism for large slippage involves a lubricating gas layer of thickness $e$ with
viscosity $\eta_g$ much smaller than that of the liquid, the so-called `gas cushion model'\cite{vinogradova.oi:1995a} \begin{equation}
b=e\left(\frac{\eta}{\eta_{\rm g}}- 1\right)\simeq e \frac{\eta}{\eta_{\rm g}}
\end{equation}
 Taking into account that $\eta/\eta_{\rm g}\approx 50,$ the variation of the SH texture height, $e$, in the typical interval $0.1-10$ $\mu$m gives $b=5-500$ $\mu$m. The composite nature of the texture requires
regions of very low slip (or no slip) in direct contact with the liquid,
so the effective slip length of the surface, $b_{\rm eff}$, is smaller than $b$. Still, one can expect that a rational design of such a texture could dramatically reduce the hydrodynamic force at relatively large O(10$\mu$m) distances.

Previous theoretical investigations of a flow past SH surfaces have addressed the questions of effective hydrodynamic~\cite{bocquet2007,feuillebois.f:2009,belyaev.av:2010a,Lauga03} and electro-osmotic~\cite{Squires08,bahga:2009} slippage. We are unaware of any previous work that has studied how the squeeze film drainage between surfaces would be modified by the occurrence of the effective slip. The main difference from the simple model of a constant slip length, Eq.~(\ref{fast}), used before for a smooth isotropic hydrophobic surfaces is that the effective slip length is itself not a characteristic of a heterogeneous (and, in general case, anisotropic) liquid/wall interface solely, but also depends on flow configuration, which in turn is determined by the smallest length scale of the problem, $H$, $b$, or roughness periodicity, $L$. Of course this immediately raises a difficulty: the decrease in $H$ during the hydrodynamic interaction with SH surfaces would inevitably modify $b_{\rm eff}$.

In this paper, we explore  what happens
when a hydrophilic disc is driven towards a SH surface in the Cassie state.
After describing the general
theory in
the following section, we present the results
and discussion of
the effect of SH slip in case of anisotropic and isotropic textures. Then follows a concluding section.

\section{General theory}

\subsection{Model}

We consider a circular hydrophilic disk of radius $R$, which is parallel to and at a small distance $H$ ($\ll R$) from a SH plane. Surfaces are immersed into viscous Newtonian liquid, and the pressure at the edge of the disk is atmospheric ($p=p_0$). The disk moves towards a plane with a constant speed $\textbf{U}$. This motion gives rise to an opposing force on the disk, which we aim to calculate.

\begin{figure}
(a) \includegraphics*[width=0.35\textwidth, trim=0 10 0 0]{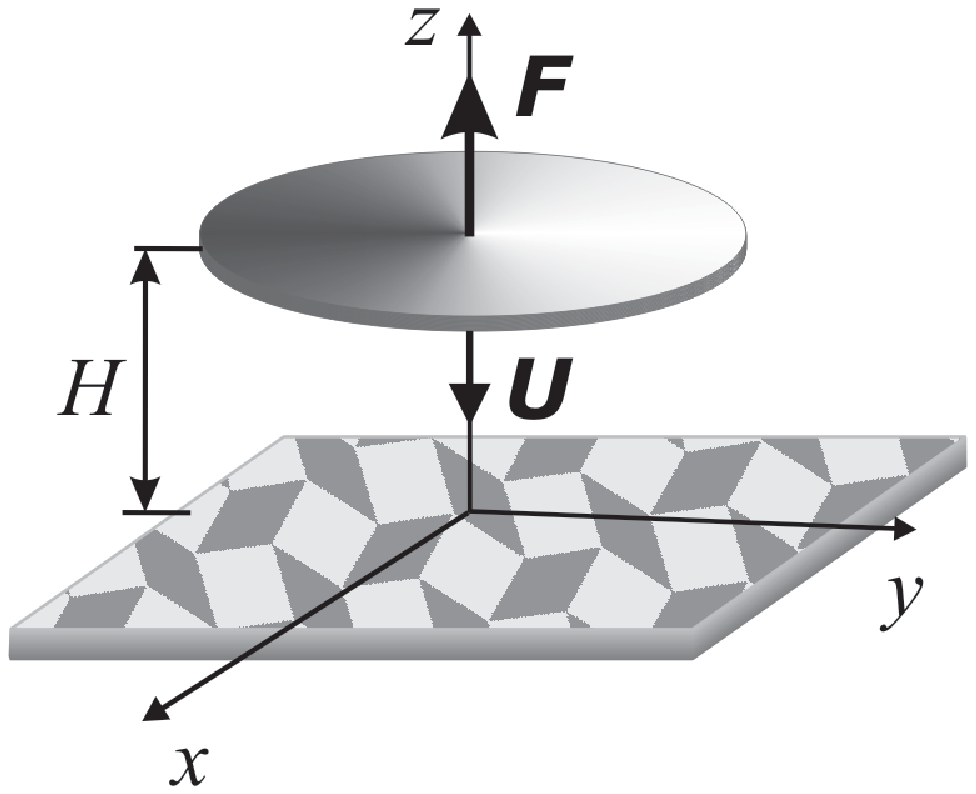}\\
(b)  \includegraphics*[width=0.3\textwidth]{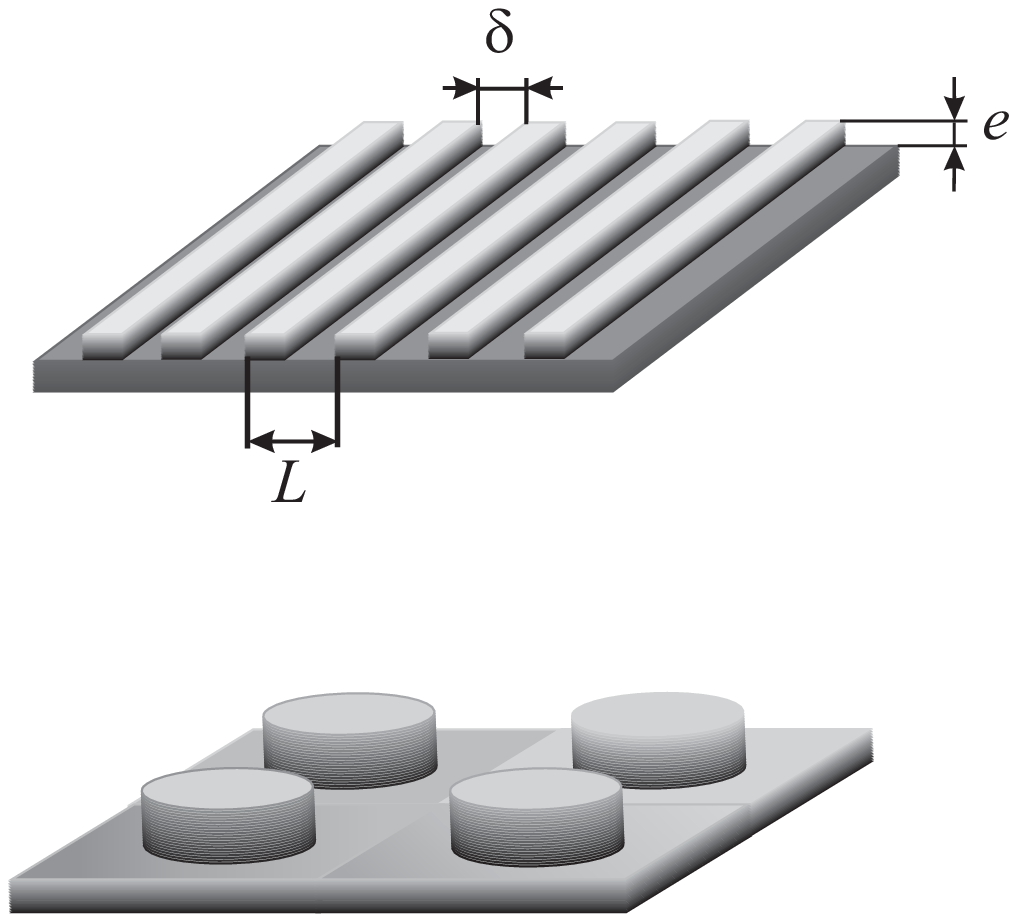}
  \caption{(a) Sketch of a hydrophilic disk approaching a superhydrophobic surface, where a  texture is represented in terms of patches of flow boundary conditions; (b) Examples of anisotropic (array of grooves) and isotropic (pillars) textures.}
  \label{fig:geometry}
\end{figure}

We examine an idealized
SH surface in the Cassie state sketched in~\ref{fig:geometry}
where a liquid slab lies on top of the surface
roughness. The liquid/gas interface is assumed to be flat with no meniscus
curvature, so that the modeled SH surface
appears as a perfectly smooth with a pattern of
boundary conditions. The latter are taken as no-slip ($b_1=0$) over solid/liquid
areas and as partial slip ($b_2=b$) over gas/liquid regions. We denote
as $\delta$ a the typical length
scale of gas/liquid areas. The fraction of solid/liquid
areas will be denoted $\phi_1=(L-\delta)/L$, and of gas/liquid area $\phi_2=1-\phi_1=\delta/L$.
Overall, the description of a SH surface we use is similar to
those considered in Refs~\cite{feuillebois.f:2009,ybert.c:2007,belyaev.av:2010a,cottin.c:2004,cottin_bizonne.c:2003.a}.
In this idealization, some assumptions may have a possible influence
on the friction properties and, therefore, a hydrodynamic force. First, by assuming flat
interface, we have neglected an additional mechanism for a dissipation connected with the meniscus curvature~\cite{harting.j:2008,lauga2009,sbragaglia.m:2007}. Second, we ignore a possible transition towards impaled (Wenzel) state that can be provoked by additional pressure in the liquid phase~\cite{pirat.c:2008,reyssat.m:2008}. Both effects are expected to modify a hydrodynamic resistance force~\cite{lecoq.n:2004,vinogradova.oi:2006,steinberger.a:2007}.

\subsection{Governing equations}

We chose a Cartesian coordinate system with the origin at the SH surface, and $z$-axis directed towards the center of the hydrophilic disk. In our case of a thin liquid film and small Reynolds numbers ($Re\ll 1$) they can be substantially simplified since the lateral component of the velocity field is large as compared with the normal component ($v_{\tau} \gg v_z $), and $(\nabla_{\tau}\textbf{v}_{\tau}) \ll \partial v_{\tau}/\partial z$.
The Navier-Stokes equations are then reduced to
\begin{equation}\label{NS}
  \eta \frac{\partial^2 \textbf{v}_{\tau}}{\partial z^2} \cong \nabla_{\tau} p, \quad   \frac{\partial p}{\partial z} \cong 0,
\end{equation}
where $p$ is pressure, $\textbf{v}_{\tau}=v_x \textbf{e}_x + v_y \textbf{e}_y$ is the lateral velocity, and $\nabla_{\tau}$ is the differential operator in plane $(x,y)$, given by
\begin{equation}\label{NablaTau}
  \nabla_{\tau}= \frac{\partial}{\partial x}\textbf{e}_x + \frac{\partial}{\partial y}\textbf{e}_y
\end{equation}
The continuity equation takes the form:
\begin{equation}\label{Cont}
  \frac{\partial v_z}{\partial z}+(\nabla_{\tau}\textbf{v}_{\tau}) = 0.
\end{equation}

At $z=H$ we have usual no-slip condition, while the boundary condition at $z=0$ reflects a tensorial hydrodynamic slip~\cite{Bazant08}
\begin{equation}\label{BC_0}
  z=0: \quad (v_{\tau})_i=b_{ij}\frac{\partial (v_{\tau})_j}{\partial z}, \; v_z=0;
\end{equation}
\begin{equation}\label{BC_H}
  z=H: \quad \textbf{v}_{\tau}=0, \; v_z=-U.
\end{equation}
Here and below we use the index form of representation for tensors and vectors: by assuming summation over a repeated index, all indices (as like $i$, $j$, $k$, etc.) can be equal to either $x$ or $y$. In particular, the first equation of (\ref{NS}) takes a form
\begin{equation}\label{NS_index}
  \eta \frac{\partial^2 (v_{\tau})_i}{\partial z^2} = \nabla_i p(x,y),
\end{equation}
where $\nabla_i\equiv(\nabla_{\tau})_i$.

\subsection{Analysis}

Expression~(\ref{NS_index}) can be integrated twice over $z$, yielding the general solution for lateral velocity components $(v_{\tau})_i$, $i={x,y}$. In the classical Reynolds problem this gives two scalar constants, which should be determined via boundary conditions. Since in general case the SH texture is anisotropic, we have to assume the tensorial character of these integration constants, and find
\begin{equation}\label{v_tau_general}
  (v_{\tau})_i = \frac{\nabla_i p}{2 \eta} \left( z^2 \delta_{ij} - A_{ij}z - B_{ij} \right).
\end{equation}
Here $\delta_{ij}$ is the Kronecker delta (two-dimensional), $A_{ij}$ and $B_{ij}$ are constant tensors, that can be determined from conditions (\ref{BC_0}) and (\ref{BC_H}). First, we find
\begin{equation}\label{diffv}
  \frac{\partial (v_{\tau})_i}{\partial z} = \frac{\nabla_i p}{2 \eta} \left( 2 z \delta_{ij} - A_{ij} \right).
\end{equation}
Then by substituting (\ref{diffv}) into (\ref{BC_0}) we get the tensorial relation
\begin{equation}\label{relation1}
  B_{ik} = b_{ij}A_{jk}.
\end{equation}
Finally, by using condition (\ref{BC_H}) together with (\ref{relation1}) we get
\begin{equation}\label{relation2}
  A_{ik} H + b_{ij}A_{jk}=H^2 \delta_{ik}.
\end{equation}
Equations (\ref{relation1}) and (\ref{relation2}) determine unknown constants in the expression for a tangential velocity, Eq.~(\ref{v_tau_general}).

To simplify further analysis, we now align basis vectors with principal directions of the slip length tensor $\{b_{ij}\}$
\begin{equation}\label{tensor_B}
   \bf{b_{\rm eff}}=\left\|\begin{array}{cc}
                b_{\rm eff}^{\parallel} & 0 \\
                0 & b_{\rm eff}^{\perp}
              \end{array}\right\|,
\end{equation}
where the eigenvalues $b_{\rm eff}^{\parallel}$ and $b_{\rm eff}^{\perp}$ are effective slip lengths in the fastest and slowest directions, correspondingly. These values can be related to the components of the effective channel permeability tensor, that determines the average fluid flux across the channel's cross-section, and, thus, depend on the gap thickness $H$.

Now we can explicitly calculate components of $\{A_{ij}\}$ and $\{B_{ij}\}$ and conclude that their principal directions coincide with those of $\{b_{ij}\}$ tensor
\begin{equation}\label{compA}
  A_{xx}=\frac{H^2}{H+b_{\rm eff}^{\parallel}}; \quad A_{xy}=A_{yx}=0; \quad A_{yy}=\frac{H^2}{H+b_{\rm eff}^{\perp}};
\end{equation}
\begin{equation}\label{compB}
  B_{xx}=b_{\rm eff}^{\parallel}\frac{H^2}{H+b_{\rm eff}^{\parallel}}; \quad B_{xy}=B_{yx}=0; \quad B_{yy}=b_{\rm eff}^{\perp}\frac{H^2}{H+b_{\rm eff}^{\perp}}.
\end{equation}

By integrating the continuity equation (\ref{Cont})
\begin{equation}
  U=\int\limits_0^H{(\nabla_{\tau}\textbf{v}_{\tau})dz},
\end{equation}
we obtain the expression for the relative speed of surfaces
\begin{equation}\label{velU}
  U=\frac{H^3}{6\eta}\nabla_i \nabla_i p - \frac{H^2}{4\eta}\nabla_i \left(A_{ij} \nabla_j p \right) - \frac{H}{2\eta} \nabla_i \left(B_{ij} \nabla_j p \right),
\end{equation}
which represents a partial differential equation for pressure:
\begin{equation}\label{PDEpressure}
  C_x\frac{\partial^2 p}{\partial x^2} + C_y\frac{\partial^2 p}{\partial y^2} = - U,
\end{equation}
where
\begin{equation}
C_x=\frac{H^3}{12\eta}\frac{H+4 b_{\rm eff}^{\parallel}}{H+b_{\rm eff}^{\parallel}}
 \end{equation}
 and
 \begin{equation}
 C_x=\frac{H^3}{12\eta}\frac{H+4 b_{\rm eff}^{\perp}}{H+b_{\rm eff}^{\perp}}
 \end{equation}

The exact solution of this partial differential equation satisfying the boundary condition for pressure at the edge of disk is
\begin{equation}\label{pressure}
 p=p_0 + \frac{U}{2} \frac{(R^2-r^2)}{(C_x+C_y)},\quad r^2=x^2+y^2,
\end{equation}

The hydrodynamic resistance force $\textbf{F}$ acting on the hydrophilic disk of radius $R$ is opposite to the force exerted by the SH surface. We remark and stress that although the anisotropy of a texture leads to the reduction of the physical symmetry of the whole system, the resulting force is still directed along the axis of highest symmetry, $\textbf{e}_z$. Thus, lateral force components $F_x=F_y$ vanish due to the presence of mirror planes parallel to the $z$-axis. The only remaining  component of the drag force is denoted below as $F$, and in the first-order approximation may be evaluated as the integral over the disk's surface
\begin{equation}\label{force}
  F=  2\pi \int\limits_0^R{\left(p-p_0 -2\eta\frac{d v_z}{dz}\right) r\:dr}
\end{equation}
However, it first-order approximation we may omit the last term in the integrand, and obtain~\cite{note1}
\begin{equation}\label{force1}
  F= \frac{3}{2}\frac{\pi\eta U R^4}{H^3} f^{\ast}_{\rm eff}, 
\end{equation}
where the correction for an effective slip is
\begin{equation}\label{force2}
  f^{\ast}_{\rm eff}=\frac{F}{F_R}= 2\left[\frac{H+4 b_{\rm eff}^{\parallel}(H)}{H+b_{\rm eff}^{\parallel}(H)}+\frac{H+4 b_{\rm eff}^{\perp}(H)}{H+ b_{\rm eff}^{\perp}(H)}\right]^{-1}.
\end{equation}
Thus the effective correction for a SH slip is the harmonic mean
of corrections expressed through effective slip lengths in two principal directions,

\begin{equation}\label{fast_sum}
    \frac{1}{f^{\ast}_{\rm eff}}=\frac{1}{2}\left(\frac{1}{f^{\ast,\parallel}_{\rm eff}}+\frac{1}{f^{\ast,\perp}_{\rm eff}}\right)
\end{equation}
In case of an isotropic textures, all directions are equivalent with $b_{\rm eff}^{\parallel}=b_{\rm eff}^{\perp}=b_{\rm eff}$, so we get
\begin{equation}\label{force3}
 f^{\ast}_{\rm eff}= \frac{F}{F_R}= \frac{H+b_{\rm eff}(H)}{H+4 b_{\rm eff}(H)}
\end{equation}
Note a similarity to Eq.~(\ref{fast}). The only difference is that slip length is effective and dependent on $H$. Obviously, the case $b_{\rm eff}^{\parallel}=b_{\rm eff}^{\perp}=0$ corresponds to $f^{\ast}_{\rm eff}=1$ and gives Eq.~(\ref{F0})

\section{Results and discussion}

In the preceding section, we derived a general expression for $f^{\ast}_{\rm eff}$, which relates it to the effective slip length of the SH wall and the gap. In order to quantify the reduction of a drag force due to a presence of SH wall, this expression will now be examined for some specific anisotropic and isotropic textures, where analytical or numerical expressions for $b_{\rm eff}^{\parallel, \perp}$ have been obtained.

\begin{figure}
  (a)\includegraphics*[width=0.17\textwidth]{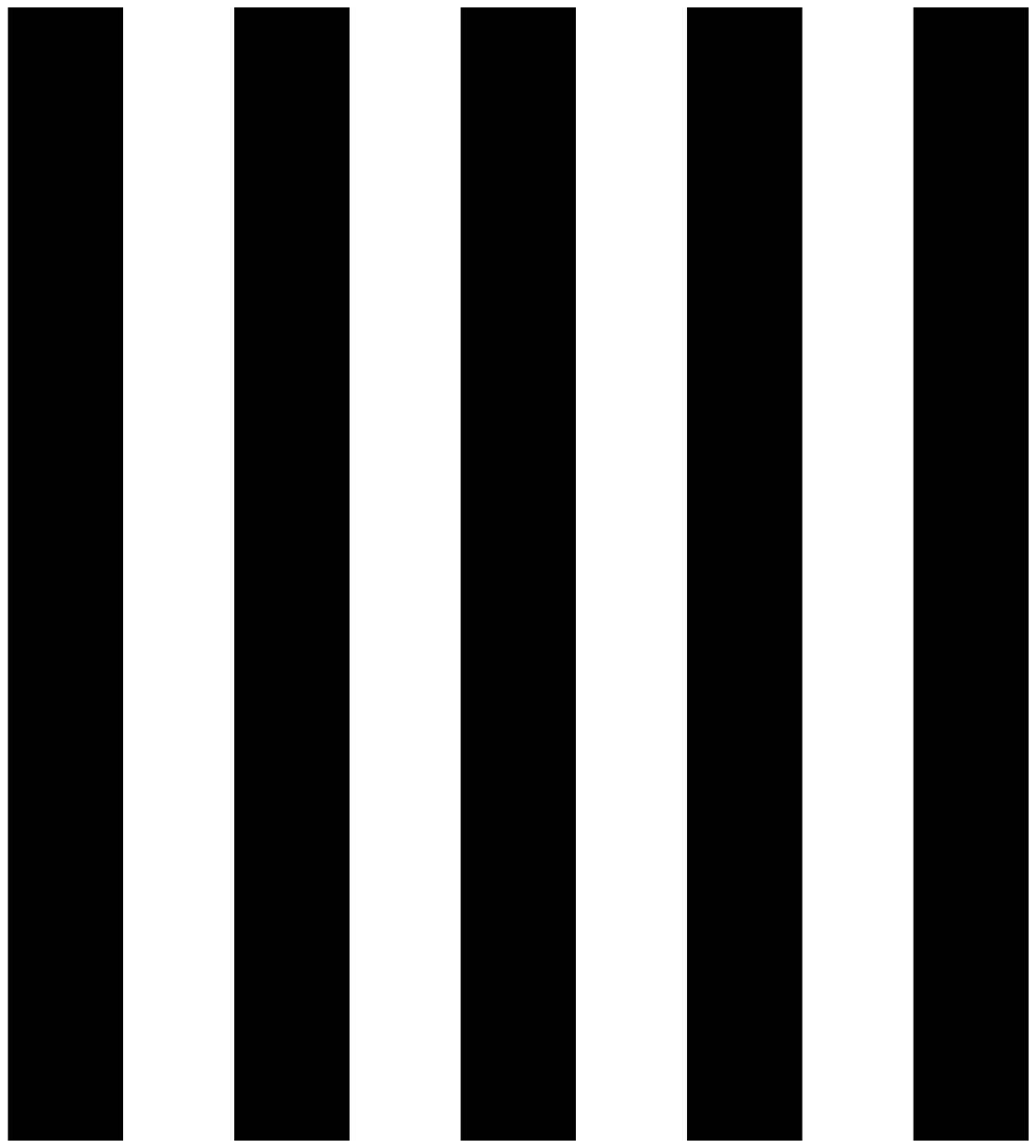}
(b)\includegraphics*[width=0.17\textwidth]{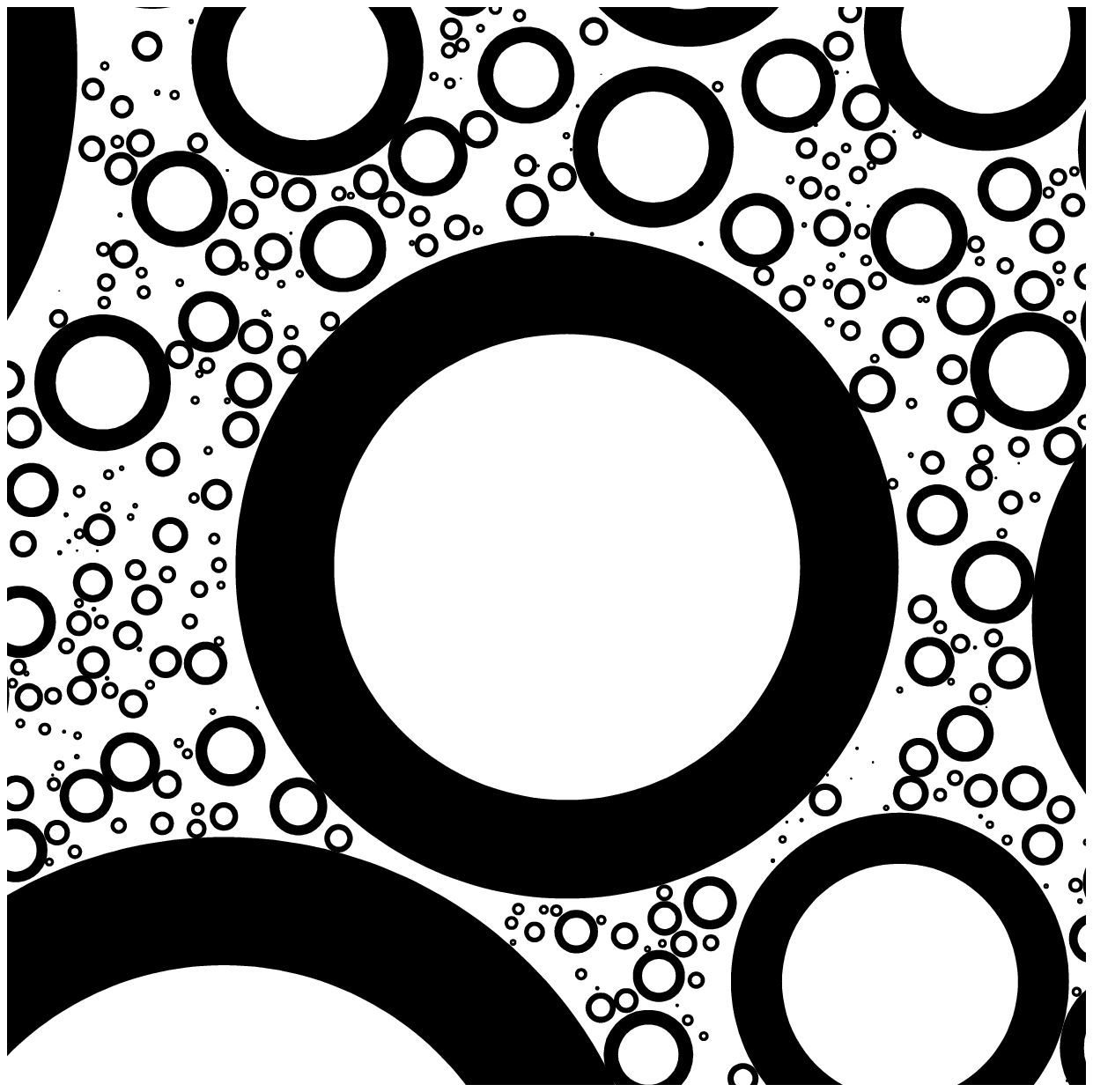}\\
(c)\includegraphics*[width=0.17\textwidth]{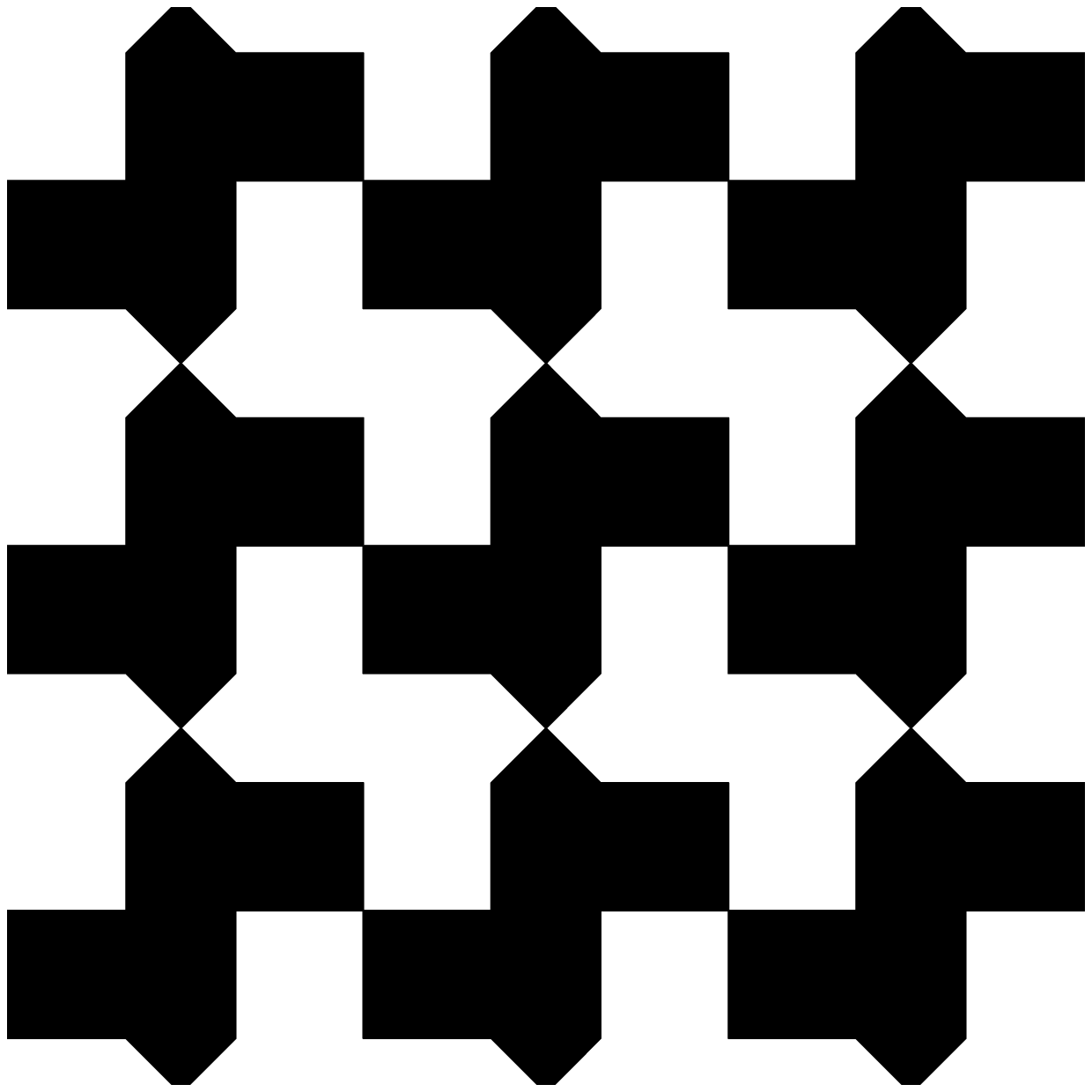}
(d)\includegraphics*[width=0.17\textwidth]{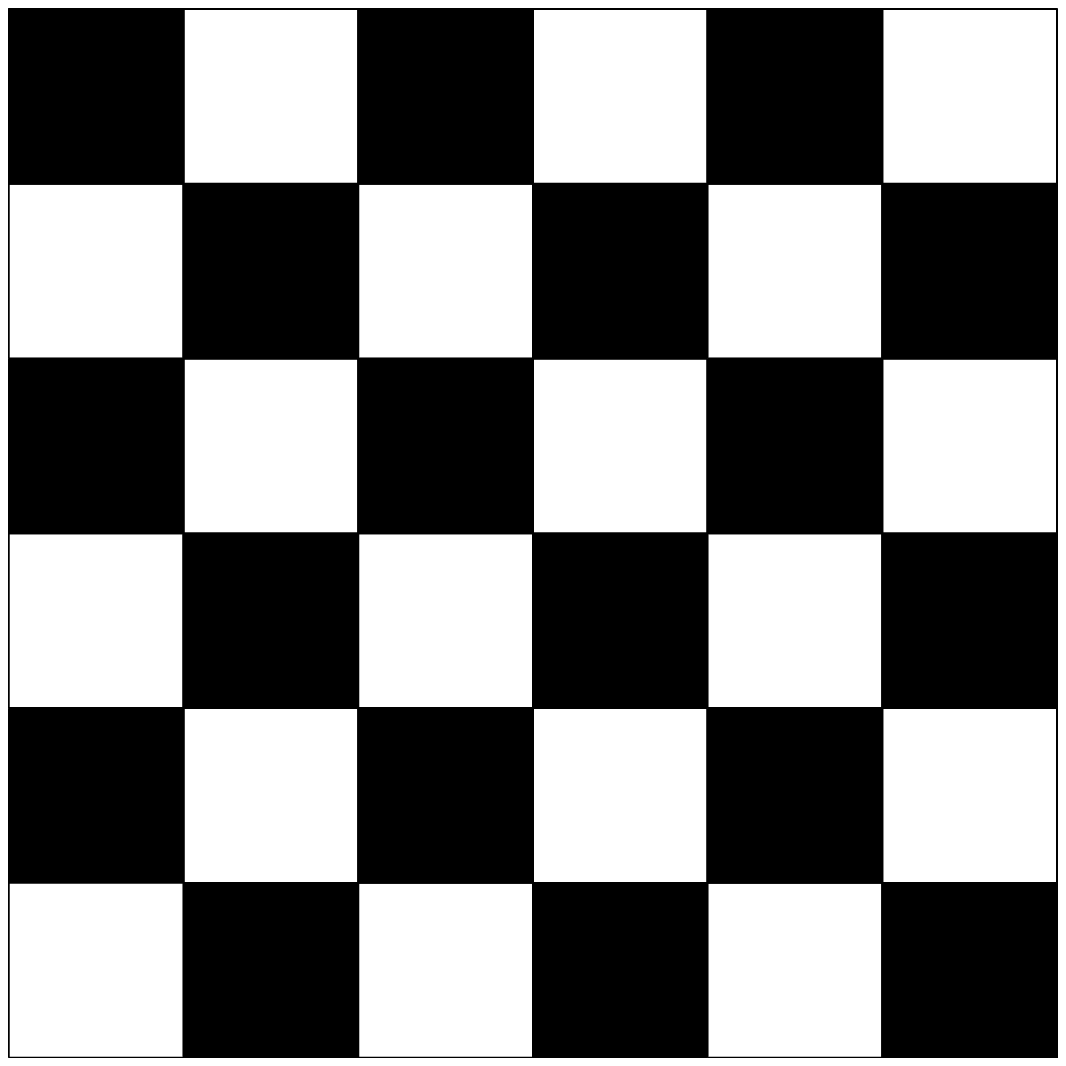}
  \caption{Special textures arising in the theory: (a)
  stripes, which attain the Wiener bounds of maximal and
  minimal effective slip, if oriented parallel or perpendicular to the
  pressure gradient, respectively; (b) the Hashin-Shtrikman
  fractal pattern of nested circles, which attains the maximal/minimal slip among all
  isotropic textures (patched should fill up the whole space, but their number is limited here for clarity); and (c) the Schulgasser and (d) chessboard textures, whose
  effective slip follows from the phase-interchange
  theorem.}
  \label{fig:designs}
\end{figure}

\subsection{Anisotropic textures}

 To highlight effects of anisotropy and to explore the effect of gap on the force, we now focus on flat, periodic, striped  super-hydrophobic surface (sketched in~\ref{fig:designs}(a)), where the local (scalar) slip length $b$ varies only in one direction. Such surfaces have been already used for reduction in pressure-driven flows~\cite{ou2005} and enhancement of mixing~\cite{ou.j:2007}.

 The problem of flow past stripes has also been examined theoretically. Effective slip lengths $b_{\rm eff}^{\parallel, \perp} (H)$ in case of an arbitrary channel thickness can be calculated semi-analytically following approach~\cite{sbragaglia.m:2007}, and the details will be published elsewhere.  In the case of thick channel ($H \gg L$) the effective hydrodynamic slip reads~\cite{belyaev.av:2010a}
\begin{equation}\label{beff_par_largeH}
  b_{\rm eff}^{\parallel} = \frac{L}{\pi} \frac{\ln\left[\sec\left(\displaystyle\frac{\pi \phi_2}{2 }\right)\right]}{1+\displaystyle\frac{L}{\pi b}\ln\left[\sec\displaystyle\left(\frac{\pi \phi_2}{2 }\right)+\tan\displaystyle\left(\frac{\pi \phi_2}{2}\right)\right]},
\end{equation}
\begin{equation}\label{beff_ort_largeH}
  b_{\rm eff}^{\perp} = \frac{L}{2 \pi} \frac{\ln\left[\sec\left(\displaystyle\frac{\pi \phi_2}{2 }\right)\right]}{1+\displaystyle\frac{L}{2 \pi b}\ln\left[\sec\displaystyle\left(\frac{\pi \phi_2}{2 }\right)+\tan\displaystyle\left(\frac{\pi \phi_2}{2}\right)\right]}.
\end{equation}
Flow in a large channel does not depend on $H$, and is controlled by the ratio of the local slip length $b$ to texture period $L$. At $b/L \gg 1$ expressions ~(\ref{beff_par_largeH})-(\ref{beff_ort_largeH}) turn to
\begin{equation}\label{effectiveslip}
b_{\rm eff}^{\parallel} = \frac{L}{\pi }\Xlog\Big[\Xsec\left(\frac{\pi \phi_2}{2 } \right)\Big]
\ \ \ \mbox{ and } \ \ \   b_{\rm eff}^\perp = \frac{b_{\rm eff}^\|}{2}
\end{equation}
suggested earlier for a perfect ($b=\infty$) local slip\cite{philip.jr:1972,Lauga03,cottin.c:2004,sbragaglia.m:2007,bahga:2009}. At $b/L \ll 1$  Eqs.~(\ref{beff_par_largeH})-(\ref{beff_ort_largeH})
predict a simple surface average isotropic effective slip
\begin{equation}\label{beff_smallH_limit1}
   b_{\rm eff}^{\perp, \parallel} \simeq  b\phi_2.
\end{equation}
  In the case of thin channels ($H \ll L$) striped surfaces were shown to provide rigorous upper and lower Wiener bounds on the effective slip over all possible two-phase patterns~\cite{feuillebois.f:2009}
 \begin{equation}\label{b_stripes}
    b_{\rm eff}^{\parallel} = \frac{b H \phi_2}{H + b \phi_1},\quad b_{\rm eff}^{\perp} = \frac{b H \phi_2}{H + 4b \phi_1}
\end{equation}
At $b/H\gg 1$ these give truly tensorial anisotropic effective slip
\begin{equation}\label{beff_smallH_limit2}
  b_{\rm eff}^{\parallel} = H \frac{\phi_2}{\phi_1},\quad b_{\rm eff}^{\perp} = \frac{b_{\rm eff}^{\parallel}}{4},
\end{equation}
but at  $b/H\ll 1$ it
leads to Eq.~(\ref{beff_smallH_limit1}).
   The above formulae well illustrate the fact that effective boundary conditions are controlled by the smallest length scale of the problem.

\begin{figure}
 \includegraphics [width=7.5 cm]{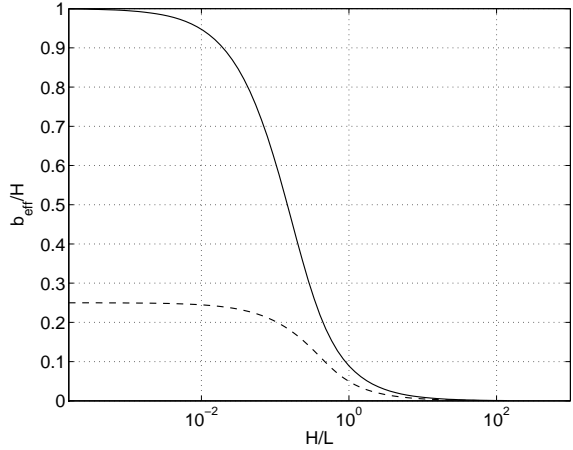}
  \caption{Eigenvalues $b_{\parallel}/H$ (solid curve) and $b_{\perp}/H$ (dashed curve) of the normalized slip length tensor for stick-slip stripes of period $L$ with local slip length at liquid-gas interface $b/L=1$ and slipping area fraction $\phi_2=0.5$ as a function of $H/L$. }
  \label{fig:b_eff}
\end{figure}

Thus, the effective slip, $b_{\rm eff}$, is large at $H/L \gg 1$, and decreases when $H/L$ is small. However, according to Eq.~(\ref{force2}) to reduce a drag force we need to maximize the ratio $b_{\rm eff}/H$, but not the absolute values of effective slip itself. The computed results for $b_{\rm eff}^{\parallel, \perp}/H$ presented in ~\ref{fig:b_eff} for $b/L=1$ and $\phi_2=0.5$ show that these values become discernible when $H/L=O(1)$ and smaller, by giving asymptotic values predicted by Eqs.~(\ref{beff_smallH_limit2}). They however vanish at large $H/L$, which is the consequence of the fact that according Eqs.~(\ref{beff_par_largeH})-(\ref{beff_ort_largeH}) the effective slip length, in this geometry, is mainly fixed by the texture period, $L$, so that by slightly modifying the results of~\cite{ybert.c:2007} we get $b_{\rm eff}/H \propto - (L/H) \ln (1-\phi_2)$. This suggests that for feasible ($\phi_2$ below 0.99) surfaces a significant reduction of hydrodynamic drag would be possible to obtain only at a thin gap limit.

\begin{figure}
  \includegraphics [width=7.5 cm]{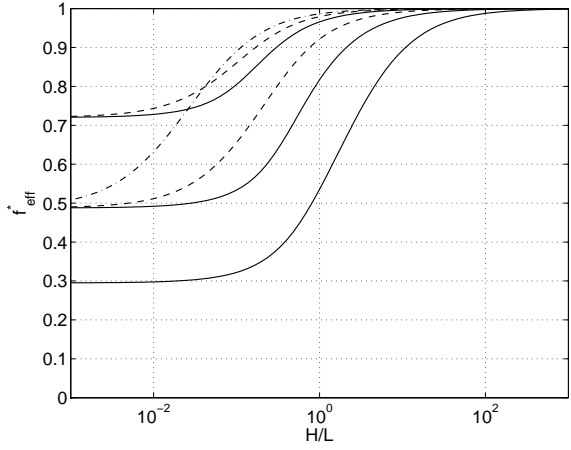}
  \caption{The correction for effective slip, $f^*_{\rm eff}$, versus dimensionless gap width $H/L$ for striped superhydrophobic surface. Solid curves correspond to local slip length $b/L=10$ (from top to bottom $\phi_2=0.2$, $0.5$ and $0.9$), dashed curves -- to $b/L=0.1$ (from top to bottom $\phi_2=0.2$ and $0.5$), dash-dotted curve - to $b/L=0.01$ and $\phi_2=0.5$. }
  \label{fig:force}
\end{figure}

This is illustrated in~\ref{fig:force}, where other calculations of $b_{\rm eff}/H$, made for several $\phi_2$ and $b/L$, were used to compute the correction for effective slip, $f^{\ast}_{\rm eff}$, as a function of the gap. Indeed, at large distances all curves converge to $f^{\ast}_{\rm eff}=1$. In other words, the drag force is the same that it would be in case of a hydrophilic surface, $F=F_R$. This conclusion can be derived directly from Eq.~(\ref{force2}) and is valid for any, however large, $b$. Results presented in~\ref{fig:force} show that for a thin gap the correction for effective slip has a tendency to decrease with $\phi_2$. Substitution of Eqs.~(\ref{b_stripes}) into Eq.~(\ref{force2}) allows to quantify this important result
\begin{equation}
 f^{\ast}_{\rm eff} = \displaystyle\frac{2 (H + 4 b - 3 b \phi_2) (H + b)}{2 H^2 + 10 b H + 8 b^2 + 9 b^2 \phi_2 - 9 b^2 \phi_2^2}
\end{equation}
In case of a small local slip, $b/H \ll 1$, we derive
\begin{equation}\label{case2}
  f^{\ast}_{\rm eff} \simeq  1- 3 \frac{b}{H}\phi_2
  \end{equation}
However, a more important limit that would represent a minimal possible, but feasible, value of $f^{\ast}_{\rm eff}$ for a  striped texture of a given $\phi_2$ can be attained in case of a large local slip, $b/H \gg 1$
\begin{equation}
 f^{\ast}_{\rm eff} \simeq \frac{2(4-3\phi_2)}{8+9\phi_2-9\phi_2^2}
\end{equation}
This expression, in particular, shows that $ f^{\ast}_{\rm eff}$ varies in the interval from 1 to 1/4 for $\phi_2$ between 0 and 1, which is in agreement with initial expectations.

\subsection{Isotropic textures}

Consider now isotropic structures,
without a preferred direction. Textures such as arrays of pillars (posts) or hollows represent a very important experimental geometry~\cite{joseph.p:2006,choi.ch:2006b,steinberger.a:2007}. No exact analytical or semi-analytical solution of the Stokes equations has been performed up to now for isotropic patterns for a gap of arbitrary thickness or even in the limit of thick channel. Based on numerical results~\cite{ybert.c:2007} we conclude that  all these textures provide effective slip confined between expected for transverse and longitudinal stripes except as in the limit of vanishing solid area ($\phi_2 \to 1$). In the latter case, scaling arguments~\cite{bocquet2007} suggested for patterns of individual pillars, where the largest effective slip is expected, $b_{\rm eff}/H \propto (L/H) /(\pi  \sqrt{1-\phi_2})$. Obviously, with the realistic $\phi_2$ this cannot change the above conclusion made for anisotropic surfaces, that a significant reduction of hydrodynamic drag would be possible only for a thin gap. Therefore, below we focus on a thin gap situation, by trying to highlight the effect of texture geometry.

\begin{figure}
  \includegraphics [width=7.5 cm]{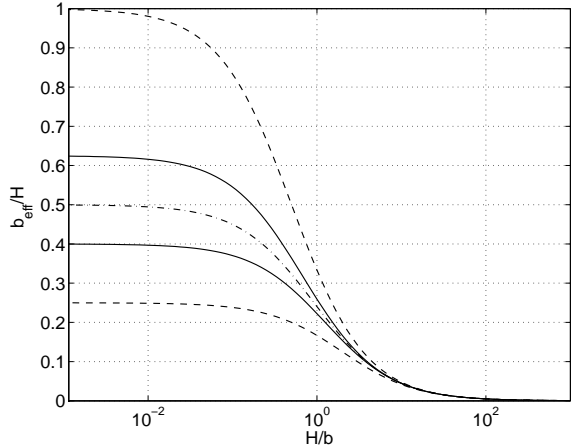}
     \caption{Effective slip length, $b_{\rm eff}/H$, versus $b/H$ [for $\phi_2=0.5$] in a thin gap limit, $H\ll L$. Superhydrophobic surfaces are: anisotropic stick-slip stripes attaining Wiener bounds (dashed curves), isotropic textures attaining Hashin-Shtrickman bounds (solid curves) and satisfying the phase-interchange theorem (dash-dotted curve). }
  \label{fig:b}
\end{figure}

Rigorous (Hashin-Shtrikman) upper and lower bounds on an effective slip length for arbitrary isotropic textures, given only the area fraction and local slip lengths, where calculated analytically in~\cite{feuillebois.f:2009,feuillebois.f:2010}.
These bounds can be attained by the special Hashin-Shtrikman fractal pattern~\cite{Torquato:2002} sketched in~\ref{fig:designs}(b). For one bound, space is filled by disks of all sizes, each containing a circular core of one component and a thick ring of the other (with proportions set by the concentration), and switching the components gives the other bound. Fractal geometry is not necessary, however, since periodic honeycomb-like structures
can also attain the bounds~\cite{Torquato-Gibiansky-Silva-Gibson:1998}.
For a situation considered here the corresponding upper bound for the effective slip length in case of isotropic surfaces can be deduced from the general result~\cite{feuillebois.f:2009,feuillebois.f:2010} as
\begin{equation}\label{b_HSU}
    b_{\rm eff} = \displaystyle \frac{b H \phi_2 (2 H + 5 b)}{H (2 H + 5 b) + b \phi_1 (5 H + 8b)}
    \end{equation}
    The lower bound is then
    \begin{equation}\label{b_HSL}
b_{\rm eff} = \frac{2 b H \phi_2}{2 H + 5 b \phi_1}.
\end{equation}
At large $b/H$ these give for upper and lower bounds

\begin{equation}\label{b_HSUl}
    b_{\rm eff} = \displaystyle \frac{5 H \phi_2}{8 \phi_1}, \quad {\rm and} \quad b_{\rm eff} = \displaystyle \frac{2 H \phi_2}{5 \phi_1},
    \end{equation}
correspondingly, i.e. similarly to anisotropic stripes (cf. Eq.~\ref{beff_smallH_limit2}), $b_{\rm eff}/H$ scales as $\propto \phi_2/\phi_1$.
The bounds for $b_{\rm eff}/H$ are plotted versus  $H/b$ in~\ref{fig:b}. Also included are results for Wiener bounds, Eqs.~(\ref{b_stripes}), plotted in a same way. Finally, for completeness we add phase interchange results, which in  the
particular case of a medium which is invariant by a $\pi/2$
rotation followed by a phase interchange, gives
\begin{equation}
  b_{\rm eff}= \frac{3 H}{\displaystyle 4-\sqrt{1+\frac{3 b}{H+b}}} -H,
\label{phase_interchange}
\end{equation}
Obviously, $\phi_1=\phi_2=0.5$ for such a medium. Classical
examples of such an isotropic texture are the Schulgasser proposal, or a family of chessboards, examples are shown in~\ref{fig:designs}(c) and (d). If $b/H \gg 1$ for textures that follow phase interchange theorem we simply get
\begin{equation}\label{Sl}
  b_{\rm eff}= \frac{ H}{2}
\end{equation}
~\ref{fig:b} shows that Hashin-Shtrikman bounds
are relatively close  and confined between Wiener ones.   It can also be seen that all curves behave similarly, by vanishing at large $H/b$. This is a consequence of the fact that in this limit the effective slip coincides with the average, $b_{\rm eff} \simeq \phi_2 b$ (cf Eq.~(\ref{beff_smallH_limit1})). At small $H/b$ all plotted curves give a plateau described by Eqs.~(\ref{beff_smallH_limit2}), (\ref{b_HSUl}), and (\ref{Sl}), depending on the texture. Its height is controlled solely by $\phi_2/\phi_1$ and texture type.

Using Eq.~(\ref{force3}) together with Eqs.~(\ref{b_HSU}), (\ref{b_HSL}) we obtain the corresponding lower
\begin{equation}\label{HS_upper}
    f^{\ast}_{\rm eff} = \displaystyle \frac{(H+b)(8b-3\phi_2 b +2H)}{(H+4b)(2b+3\phi_2 b+2H)}
\end{equation}
and upper
\begin{equation}\label{HS_lower}
    f^{\ast}_{\rm eff} = \displaystyle \frac{2H+5b-3\phi_2 b}{2H+5b+3\phi_2 b}
\end{equation}
bounds for a correction for effective slip. We remark here, that the lower bound for $f^{\ast}_{\rm eff}$ corresponds to the upper bound for $ b_{\rm eff}$, and \emph{vice versa}.
Similarly, by combining Eq.~(\ref{force3}) together with phase interchange results, Eqs.~(\ref{phase_interchange}), we derive
\begin{equation}
  f^{\ast}_{\rm eff}=\displaystyle\sqrt{ \frac{H+b}{H+4 b}}
\label{phase_interchange2}
\end{equation}

\begin{figure}
 \includegraphics [width=7.5 cm]{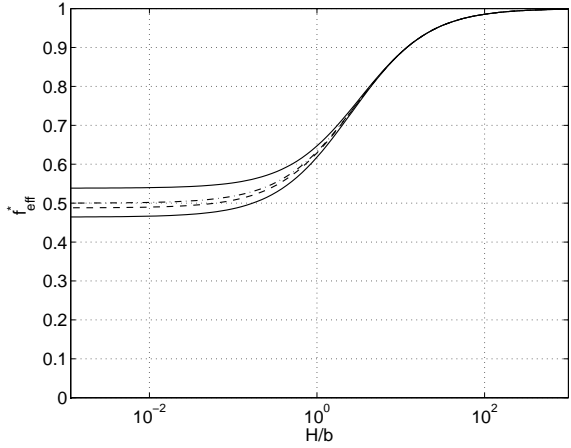}
   \caption{The correction for effective slip, $f^*_{\rm eff}$, versus $b/H$ [for $\phi_2=0.5$] in a thin gap limit, $H\ll L$. Superhydrophobic surfaces are: anisotropic stick-slip stripes attaining Wiener bounds (dashed curve), isotropic textures attaining Hashin-Shtrickman bounds (solid curves) and satisfying the phase-interchange theorem (dash-dotted). }
  \label{fig:fast1}
\end{figure}

The Hashin-Strickman bounds for $f^{\ast}_{\rm eff}$ are plotted in~\ref{fig:fast1}.  The results for textures satisfying phase interchange theorems are also included in~\ref{fig:fast1}, and confined between Hashin-Strickman bounds as predicted by the theory. To examine the significance of isotropy/anisotropy more
closely, the short-distance region of one of the curve from~\ref{fig:force} is reproduced in~\ref{fig:fast1} in the corresponding coordinates. It turns out that the results for stripes are also confined between Hashin-Strickman bounds for $f^{\ast}_{\rm eff}$, so that in general case anisotropy would not help to reduce/enhance a drag force. We stress however, that the Hashin-Strickman bounds
are fairly close, so the theory provides a good sense of the possible $f^{\ast}_{\rm eff}$ of any
isotropic or even anisotropic texture. The data presented in~\ref{fig:fast1} show that at very large distances, the resistance to approach flow is the same as
it would be in the Reynolds problem, with no slippage on both surfaces. A straightforward calculation shows that at small $b/H$ the useful approximation for $f^{\ast}_{\rm eff}$ for all textures would be Eq.~(\ref{case2}). This  universal behavior is confirmed by coincidence of all curves presented in~\ref{fig:fast1} in this limit. If the gap is much smaller than local slip length at the gas area, the correction for effective slip becomes smaller and turns asymptotically to constant values. For the Hashin-Strickman bounds these can be evaluated as
\begin{equation}
  f^{\ast}_{\rm eff}\simeq  \frac{8-3\phi_2}{4 (2+3\phi_2)},\quad  f^{\ast}_{\rm eff} \simeq  \frac{5-3\phi_2}{5+3\phi_2}
\label{HS_fstar_asym}
\end{equation}
Correspondingly, for a Schulgasser (or chessboard) textures  $f^{\ast}_{\rm eff} \simeq 1/2$ in this limit, which is well seen in~\ref{fig:fast1}.

\begin{figure}
 \includegraphics [width=7.5 cm]{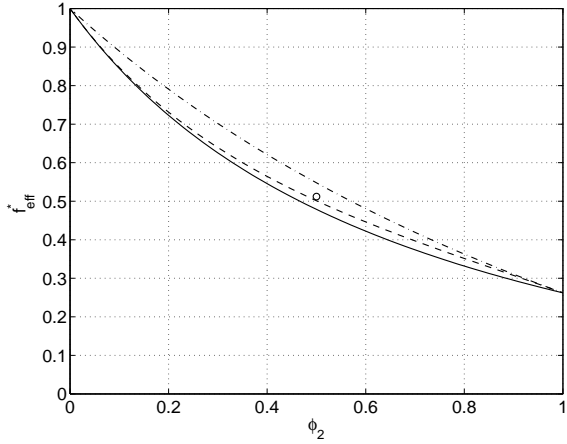}
  \caption{The correction for effective slip, $f^*_{\rm eff}$, versus  $\phi_2$ [for $b/H=15$] in a thin gap limit, $H\ll L$. Superhydrophobic surfaces are: anisotropic stick-slip stripes attaining Wiener bounds (dashed curve), isotropic textures attaining Hashin-Shtrickman bounds (solid curve for the upper bound and dash-dotted for the lower one ). The value for the chessboard or
the isotropic Schulgasser structure is also shown as a circle. }
  \label{fig:fast2}
\end{figure}

These results suggest that the key parameter determining reduction of drag is the
area fraction of gas, $\phi_2$, in contact with the liquid. This is illustrated in~\ref{fig:fast2}, where (using a relatively large $b/H$)  Hashin-Shtricknann bounds for $f^{\ast}_{\rm eff}$ are plotted versus the
liquid-gas area fraction $\phi_2$. If this is very small (or $\phi_1 \to 1$) for all
textures, the correction for slip tends to its absolute maximum, $f^{\ast}_{\rm eff} = 1$.
In the most interesting limit, $\phi_2 \to 1$, we can achieve the minimum possible value of correction for effective slip, $f^{\ast}_{\rm eff} = 1/4$, provided $b/H$ is large enough. We also notice that at small $\phi_2$ the results for stripes coincides with the lower Hashin-Shtricknann bound for $f^{\ast}_{\rm eff}$. In contrast, while for large $\phi_2$ they reduce a drag as it would be in the upper Hashin-Shtricknann bound for $f^{\ast}_{\rm eff}$.

\section{Conclusion}

We have analyzed the squeeze-film
drainage of a liquid confined between a hydrophilic disk and patterned SH surface of non-uniform
local slip length, and  have obtained
general solutions to arbitrary gap, and slip
variation. We have shown
that the decrease in the hydrodynamic force in the presence of a patterned slipping surface can be described in terms of a correction for slippage to the Reynolds formula, formulated as a function of the slip lengths in the fastest and slowest direction. Provided the separation is small compared to texture period, this correction to slippage becomes small as compared with unity, providing the significant decrease in hydrodynamic drag. We have concluded
that in all situations, to achieve a large reduction of a drag force optimizing the pattern geometry is not nearly as
important as to maximizing local slip at the gas area and the fraction of the gas phase.

\section*{Acknowledgement}

This research was partly supported by the DFG under the Priority programme ``Micro and nanofluidics'' (grant Vi 243/1-3) and by the RAS under the
Priority Program ``Assembly and Investigation of Macromolecular Structures of New Generations''.

\bibliographystyle{rsc}
\bibliography{sm_Belyaev}

\begin{mcitethebibliography}{42}
\bibitem[not()]{note1}
Note that in the derivation we neglected a term $\rho \partial
  \textbf{v}_{\tau} / \partial t$. It is straightforward to show that in our
  model, where the effective slip, $b_{\rm eff}$, depends on the gap, $H$, this
  term can also safely be omitted. This is justified provided it is small as
  compared to $ \eta v_{\tau} /H^2$, which is equivalent to a standard
  assumption of the classical Reynolds theory, Eq.~(\ref{F0}), $U \rho H /\eta
  \ll 1$\relax
\mciteBstWouldAddEndPuncttrue
\mciteSetBstMidEndSepPunct{\mcitedefaultmidpunct}
{\mcitedefaultendpunct}{\mcitedefaultseppunct}\relax
\EndOfBibitem
\end{mcitethebibliography}

\end{document}